\documentclass[review,1p,authoryear,12pt]{elsarticle}

 \usepackage{graphics}
 \usepackage{graphicx}
 \usepackage{epsfig}

\usepackage{amssymb}
 \usepackage{amsthm}

 \usepackage{lineno}

\usepackage{hyperref}
\usepackage{subfigure}
\usepackage{color}

\journal{Planetary and Space Science}
\begin{document}
\begin{frontmatter}


\title{\Large {\bf {Experimental validation of XRF inversion code for Chandrayaan-1}} \\}

\author[rvt,focal]{P.S. Athiray\corref{cor1}}
\ead{athray@gmail.com}
\author[rvt]{M. Sudhakar}
\author[cat]{M.K. Tiwari}
\author[rvt]{S. Narendranath}
\author[cat]{G. S. Lodha}
\author[cat]{S. K. Deb}
\author[iia]{P. Sreekumar}
\author[rvt]{S.K. Dash}
\cortext[cor1]{Corresponding author at:}

\address[rvt]{ISRO Satellite Centre, Vimanapura P.O., Bangalore 560 017, India}
\address[focal]{Department of Physics, University of Calicut, Thenjippalam, Kerala, India}
\address[cat]{Raja Ramanna Centre for Advanced Technology (RRCAT), Indore,
M.P., India}
\address[iia]{Indian Institute of Astrophysics (IIA), Koramangala,
Bangalore, India}

\begin{abstract}
We have developed an algorithm (x2abundance) to derive the lunar surface
chemistry from X-ray fluorescence (XRF) data for the Chandrayaan-1 X-ray
Spectrometer (C1XS) experiment. The algorithm
converts the observed XRF line fluxes to elemental
abundances with uncertainties. We validated the algorithm
in the laboratory using high Z elements (20 $<$ Z $<$
30) published in \cite{psa13}. In this paper, we complete the exercise of
validation using samples containing low Z elements, which
are also analogous to the lunar
surface composition (ie., contains major elements between 11 $<$ Z $<$ 30). The paper
summarizes results from XRF experiments performed on Lunar simulant (JSC-1A) and
anorthosite using a synchrotron beam excitation. We also discuss results
from the validation of {\it x2abundance} using Monte Carlo simulation (GEANT4 XRF
simulation).

\end{abstract}

\end{frontmatter}

\section{Introduction}
\label{intro}

Solar X-rays impinge on the lunar surface and trigger XRF emission from
elements in the upper layers (the major elements being viz., Mg, Al, Si, Ca, Ti and Fe).
Chandrayaan-1 X-ray spectrometer (C1XS) \citep{Gr09} measured this XRF
emission during several flares in the period Nov'2008 - Aug'2009. In order to
convert the observed X-ray line flux determined from C1XS spectral data, a
code `x2abundance' is
developed, where a new approach is adopted using Fundamental Parameter
method  \citep{Cb68, RB98}. A detailed description of the
algorithm of {\it x2abundance} along with the fundamental assumptions and
limitations are given in \citep{psa13} (here onwards Paper I).  In paper I, we
validated {\it x2abundance} with laboratory XRF experiments using samples
containing major elements of high atomic number (20 $<$ Z $<$ 30 - set A
samples). However, the major rock forming elements are low Z elements (Z
$<$ 26) which makes it necessary to demonstrate the validity using samples
with composition similar to the lunar surface. In this paper, we
present the validation of {\it x2abundance} using XRF experiments on lunar
analogues with elements in the range 11 $<$ Z $<$ 26. We also compare
{\it x2abundance} results with X-ray fluorescence simulation using GEANT4, a Monte Carlo
simulation tool kit.\\

A brief description of the laboratory XRF experiments on lunar analogue samples
excited with a synchrotron beam is given in Sec. (\ref{labexpt}).
Details of X-ray spectral analysis along with laboratory experimental
results and validation are given in Secs.(\ref{labana} and \ref{labres}). GEANT4 XRF simulations, results
and validation are summarized in Sec. (\ref{geant}). Finally, a short discussion on the complete work is presented in Sec. (\ref{conc}).

\section{XRF experiments}
\label{labexpt}
The objective of the current experiment is to validate {\it x2abundance} developed
for C1XS analysis using samples that are analogous to lunar surface
composition. These samples contain elements from Mg to Fe whose K-shell
X-ray emission energies ranges from ${\approx}$ 1.2 keV to ${\approx}$ 7.0 keV.

\subsection{Sample details}
Lunar surface can broadly be classified into two regions viz.,
(a) dark mare regions (high mafic content) and (b) bright highland regions (concentration of low dense minerals with low mafic content).
The mare regions are believed to be the partial melts of lunar interior erupted
on the surface. Whereas, highlands are believed to be formed during the end
stages of lunar magma ocean where the lighter mineral plagioclase feldspar
floated and crystallized on the surface. For our experiment, two samples
which closely represent the composition of lunar mare and highland region
are used. A brief description of the samples used are given here:

\begin{itemize}
	\item{Lunar Simulant (JSC-1A) : JSC-1A is a lunar mare regolith simulant
		released by NASA for research studies mined from a
		commercial cinder quarry at Merriam Crater (35$^{\circ}\mathrm{}$ 20' N, 111$^{\circ}\mathrm{}$ 17' W).
		\footnote{A volcanic cinder cone located in the San Francisco volcano field
		near Flagstaff, Arizona.} The volcanic ash deposit of Merriam
		crater is basaltic in composition, similar to the soil from Maria
		terrain of the Moon. The mined ash was processed to
		produce the simulants with mean grain size of ${\approx}$ 190
		${\mu}$m. A complete
		characterization of lunar simulant (JSC-1A) is reported by
		\citep{Ray10}.}\\

	\item{Sittampundi rock : This sample is taken from the anorthositic rocks
		available at Sittampundi (near Salem, Tamil Nadu), India,
		which are considered close to lunar highland rocks
		in composition. The plagioclase in these rock samples exhibit an
		anorthite content in the range An$_{80 - 100}$ \citep{Anb12} in
		comparison to that in Apollo returned
		samples which has an anorthite content of ${\approx}$ An$_{95}$ on average.
		Pure calcic-anorthite and labradorite
		are the dominant minerals in this sample. The sample exhibit a low mafic
		content and large concentration of Al and Ca. They are pulverized to a grain
		size of about 100 ${\mu}$m or smaller. Detailed studies on the rocks of this site
		are reported by \citep{Anb12, Sanj09}.}

\end{itemize}

Pellets of these samples were made after crushing them manually and
compressing it with a pelletizer. We assume that the pellets are a
close representation of flat and homogeneous sample.  However, in reality
the size distribution of soil particles on the Moon is not well
characterized. It is shown already that, in heterogeneous samples, the
observed XRF line strength exhibit a dependence on grain size \citep{Mar08, Nar08}, which is
assumed to be minimal for C1XS observations due to large spatial
scales \citep{Wei12}.  A detailed description of experimental
setup is given in the following section.

\subsection{Experiment details}
 As the samples contain diverse elements ranging from Mg to Fe, we
 irradiated the samples using a high intense mono-energetic synchrotron
 X-ray beam. For this purpose, we have utilized the XRF-${\mu}$ probe beam-line (BL-16) at Indus-2 synchrotron
beam facility, RRCAT, Indore, India. The Indus-2 electron storage ring is currently operating at 2.5 GeV with nominal current of 100
mA. BL-16 of Indus-2 is installed on the 5$^{\circ}\mathrm{}$ port of bending magnet and operates under high vacuum
condition of 10$^{-6}$ mbar. It is designed to work in the photon energy range
of 4-20 keV.
A double crystal monochromator (DCM) with a pair of symmetric and asymmetric Si (111) crystals (mounted
side-by-side) is placed ${\approx}$ 19m away from the source. It provides a
mono-energetic beam, which in our case is 8.0 keV, with an energy
resolution (${\delta}$E/E) of ${\approx}$ 10$^{-3}$ - 10$^{-4}$. Various salient features of BL-16 beamline and setup
are described elsewhere \citep{Ti13}. The measured energy
resolution (${\delta}$E/E) of the beamline is found to be ${\approx}$ 1.34 ${\times}$ 10$^{-4}$
at 10 keV. \\

However, it is to be noted that the X-rays from solar flares exciting the
Moon is mostly a thermal continuum along with many emission lines that
dominates in soft X-rays ($<$ 8 keV), which also exhibit spectral
variability in time.  Under such circumstances, the effect of surface
roughness and grain size coupled with observational geometry on XRF line
strength becomes important, which is not currently addressed
in this work.\\

For the current experiment, the second crystal of DCM is detuned
slightly to avoid higher order Bragg reflections from the DCM. The sample
holder is placed at 45$^{\circ}\mathrm{}$ with respect to the incident beam
direction. The monochromatic beam is allowed to pass through an
ionization chamber for monitoring incident flux (I$_o$) reaching the
sample. The fluorescent/scattered X-rays emitted from the sample are
detected using a solid state Si PIN detector (XR-100CR -  160 eV at 5.9 keV; Amptek\footnote{http://www.amptek.com/}, Inc. USA) coupled to
a digital pulse processor unit. The detector is placed in the horizontal
plane perpendicular to the incident beam direction making a phase angle
(angle between the incident beam, sample and detector) of ${\approx}$ 90$^{\circ}\mathrm{}$.
The sample to detector distance is set to ${\approx}$ 10cm. A collimated
mono-energetic X-ray beam of energy 8 keV and size ${\approx}$ 1mm ${\times}$ 1mm, is used
for the sample excitation. All XRF measurements are
done inside a vacuum chamber made out of stainless steel (SS-304) with 
pressure of ${\approx}$ 10$^{-2}$ mbar. Fig. (\ref{photo_catvac}) shows the photograph of the experimental arrangement at BL-16.\\


\section{XRF spectral analysis}
\label{labana}
Observed XRF spectra of laboratory samples are analyzed using the X-ray spectral analysis package
(XSPEC) \citep{KA66}. The best spectral fit obtained for the samples JSC-1A
and Sittampundi rock excited by an intense 8 keV mono-energetic beam are shown in
Fig. (\ref{specfitsjsc} \& \ref{specfitsano}). XRF lines of low Z elements like Al, Si, K are clearly visible in
the spectrum which are modeled as Gaussian functions. X-ray signature of
Mg is not seen in the spectrum of JSC-1A due to reduced efficiency
of the detector at lower energies around 1 keV  and also from a low probability
to excite the relatively less abundant Mg by a 8 keV beam.

\subsection{Spectral contamination and corrections}

The observed XRF spectra also exhibit lines from elements which are not present
in the samples as well as a continuum background emission. Reasons for these features
seen in the spectrum and necessary corrections applied to derive the XRF
line flux of the samples are discussed below:
\begin{itemize}
\item{XRF lines of stainless steel arising from the vacuum chamber walls are
	clearly observed (for example K-${\alpha}$ of Cr @ 5.4 keV). This is probably due to
	the interaction of uncollimated scattered 8 keV beam with the walls of
	the vacuum chamber. As some of the elements in SS-304 are also
	present in the samples
(ref. Fig. (\ref{specfitsjsc})), we have to estimate the contribution from
SS-304.
For this, we derived the XRF spectrum of SS-304 excited by an 8 keV beam
using GEANT4 Monte Carlo simulation.}
\item{Scattered continuum components of 8 keV incident beam: viz., Rayleigh (elastic) and Compton
(inelastic) are also prominently seen in the observed spectra (ref.
Fig.(\ref{specfitsjsc} \&\ref{specfitsano})). Earlier
experiments by \cite{psa13} were performed with a lower intensity X-ray
beam and hence did not reveal the presence of scattering components. To model the
scattered components, using apriori composition, we simulated both elastic
and inelastic scattered spectrum in GEANT4. We generated table models
\footnote{Table Model - ftp://legacy.gsfc.nasa.gov/caldb/docs/memos} of the
XRF spectrum of SS-304 and scatter component spectrum (elastic and
inelastic) which are compatible for XSPEC analysis. Table models used in
XSPEC are user defined models containing a grid of spectra and range of
parameter values used in the model. While fitting the data,
normalization of the differential flux is allowed to vary, retaining the
the spectral profile.}

\item{The samples also contain many trace
elements (in ppm) which are mostly high Z
(For JSC-1A ex. Ni, Zr, Sr, Ce, Ba etc.,). Many of these exhibit numerous L-
lines which lie very close to/overlap  XRF lines of major elements
(for eg: L-lines of Ba lie close to Ti \& Cr K-lines; Sr L-lines lie close to
Si K-lines) in the samples.
Presence of these trace element line features are clearly seen in the residuals in Fig.
(\ref{specfitsjsc} (a)) when the spectrum is modeled only with major
elemental lines, SS-304 lines and
scatter components.  Prominent trace element
lines fitted with Gaussian functions in the XRF spectrum are shown in Fig.
(\ref{specfitsjsc}(b) \& \ref{specfitsano}) (labeled vertically below the peaks).}

\end{itemize}

Apart from these factors, a tiny fraction of contamination is expected due
to scattering of these lines from perspex target holder. To account for
possible contribution from unobserved trace elements and scattering from perspex, XRF line
fluxes are considered to have a minimum uncertainty of 2\%. Measured XRF
line fluxes of the samples are  derived after applying corrections for
contamination. Figures \ref{specfitsjsc} \&
\ref{specfitsano} shows the best spectral fits to the
samples after including all components (walls, trace elements, scattering). Flux-fractions are computed using eq.1 from the observed flux and fed to {\it x2abundance}
along with the mono-energetic input spectrum.
\begin{equation}
	\label{ff}
	FF_{i} = \frac{F_i}{{\sum}_{i=1}^{n} F_i}
\end{equation}
\noindent
where i run across all `n' elements in the sample, $F_{i}$ -
computed X-ray line flux of $i^{th}$ element, $FF_{i}$ -  computed flux
fraction of $i^{th}$ element. Elemental abundances are derived
using {\it x2abundance} with computations performed for both cases ie., fitting
with and without trace elements and the results are discussed below.\\

\section{Laboratory experiment results}
\label{labres}
XRF lines of Mg, Na and O could not be seen
in the spectrum of JSC-1A due to the reduced effective area of the detector
below 1.2 keV. Hence the weight percentage values of these elements are
fixed to values obtained from EDX measurements (2.6\%, 1.5\% and 52.7\%
) respectively.  Table \ref{JSC1Aabundance} compares the elemental abundances of lunar simulant
(JSC-1A) derived from {\it x2abundance} along with the measured values using EDX
facility at ISAC. Column (3 \& 4) gives the abundances derived for JSC-1A
fitted with and without trace element contribution. Inclusion of trace
elements in fitting can gives abundances which are closer to the true
values.\\

For the analysis of Sittampundi sample, abundance of Oxygen is kept fixed as
57.0\% (EDX measurement). Trace elements are fitted with Gaussians based on the residuals as
a complete chemical characterization of these rocks are unavailable.
Figure (\ref{labcompplots} (b))
compares major elemental abundances derived by
{\it x2abundance} against the true abundances (EDX measurements) with
2${\sigma}$ residuals shown in the bottom panel. True elemental abundances agree well within
2${\sigma}$ errors of the derived values.\\

It should be noted that {\it x2abundance} does not demand any a priori information on the elemental weight
percentages provided all the elements are observed in the spectrum
(excluding Z $<$ 10). In this experiment, weight percentages of Mg and Na are obtained from EDX measurement, as they are not observed in the spectrum due to  sensitivity of the detector at low X-ray energies. Experiments with Silicon Drift Detectors (SDD) can overcome this problem as
it offers better quantum efficiency at low energies around ${\approx}$ 1 keV.\\

\section{GEANT4 simulation and results}
\label{geant}
GEANT4 is a Monte Carlo based toolkit to simulate the interaction of particles through matter \citep{Ago03}. It
incorporates various physics interactions, event tracking system, user-defined
geometries, digitization etc., GEANT4 covers a wide range of energies of
interaction (say 250eV to TeV energies) from optical to ${\gamma}$-rays and
charged particles. It allows the user to define the incident beam,
experimental geometry, sample composition and include necessary fundamental
physics processes.\\

We have used GEANT4 (ver 9.4) with
electromagnetic physics process including XRF and scattering (both
elastic and inelastic). We defined the simulation geometry similar to that used in
the laboratory, in order to cross validate {\it x2abundance} with
laboratory experiments. We simulated XRF emission from a set of samples listed in
Table \ref{geantsamples}. Simulated XRF line fluxes of all the elements
are assumed to have Poisson errors.
Flux-fractions are computed using eq.(1), which along with incident
spectrum serve as input to {\it x2abundance}.

A comparison plot of derived abundance vs true abundance
of all major elements for the sample JSC-1A simulated in GEANT4
experiment is shown in Fig. (\ref{geantcompplots}). It is evident from Fig.
(\ref{geantcompplots}) that the abundances derived by {\it x2abundance}
program matches well with the abundances for which GEANT4 simulation is
performed. The deviation between the two is plotted as residuals
in the bottom panel. \\

\subsection{Cross-validation}
We also tested the self-consistency of the inversion process by comparing
the measured XRF spectrum with GEANT4 simulated spectrum which used the
composition derived by {\it x2abundance} and convolved with instrument
response. Steps involved in the
cross-validation are shown in fig. (\ref{closure}). Elemental abundances of
Sittampundi sample obtained from {\it x2abundance} are provided as input for
GEANT4 XRF simulation. Fig. (\ref{closure_ff}) shows the comparison of
simulated flux-fractions plotted against the measured values. The plot also
shows the best fit to the data where the slope tending to unity
provides a good confidence in the {\it x2abundance} output. Deviation from the
expected slope of unity could be due to unobserved trace elements not
modeled in the spectrum.

\section{Conclusion}
\label{conc}
Using the established FP method, we have developed an algorithm
`{\it x2abundance}' for remote sensing XRF observations. Major
corrections required in the analysis of remote sensing
observations viz., time and energy dependent excitation source, varying
observation geometry, unknown composition, uncertain matrix effect are all
incorporated in {\it x2abundance}. It is also to be noted that {\it x2abundance}
provides the best suite elemental composition and its uncertainties.
We have validated our approach using laboratory-based XRF
experiments on metal alloys, lunar analogues and also using GEANT4 XRF
simulations. \\

With this work, we complete the process of validation of
{\it x2abundance} using laboratory XRF experiments on samples with atomic numbers (11
$<$ Z $<$ 30), which include all the major rock forming elements. We
also report the validation of the same using simulated XRF spectra. It
is shown that in all cases, the true abundances matches well within the
uncertainty of our derived abundances.\\

However, as mentioned earlier, remote sensing measurements from planetary
surfaces needs correction for the effect of surface roughness and grain
size which are to be incorporated in {\it x2abundance} in order to approach the
reality. As a future work, experiments are being planned to
address the particle size effects along with sensitive measurements of low Z
elements, whose characteristic X-ray peaks are around 1 keV.\\


\section{Acknowledgments}
The authors thank Dr. P. D. Gupta, Director RRCAT Indore for his
encouragement and support in successfully conducting the
experiment using the synchrotron facility. Our thanks to Shri.
Ajith K Singh, Shri. Ajay Khooha, Shri. Vijendra Prasad and Shri. S. R. Garg of RRCAT Indore for their timely
assistance. We also thank Ms. Uma Unnikrishnan for active discussions and
help in GEANT4 simulation using user defined samples. We thank
Prof. B.R.S. Babu and his team at the Dept., of Physics, University of
Calicut for making pellets of the samples.

\onecolumn{
\begin{figure}%
		\centering
		\includegraphics[width=15.05cm]
		{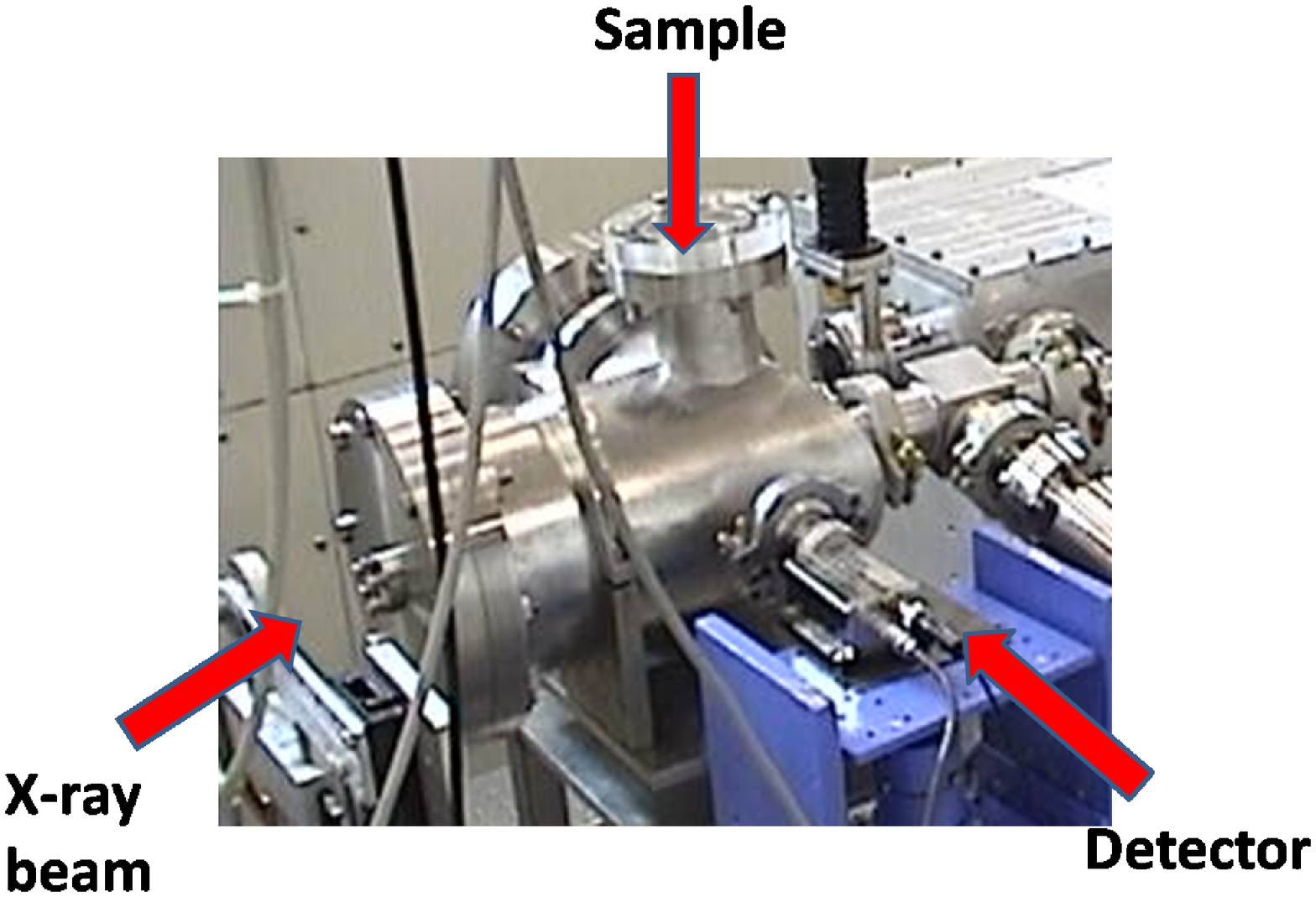}
		\caption{Photograph of the XRF experiment on lunar analogues under vacuum condition using synchrotron facility at RRCAT, Indore.}
		\label{photo_catvac}
	\end{figure}

\subfiglabelskip=0pt
\begin{figure}%
\centering
		\subfigure[XRF spectrum - Lunar simulant (JSC-1A) - Fit without trace
		elements]{
		\includegraphics[width=7.05cm,angle=-90]
		{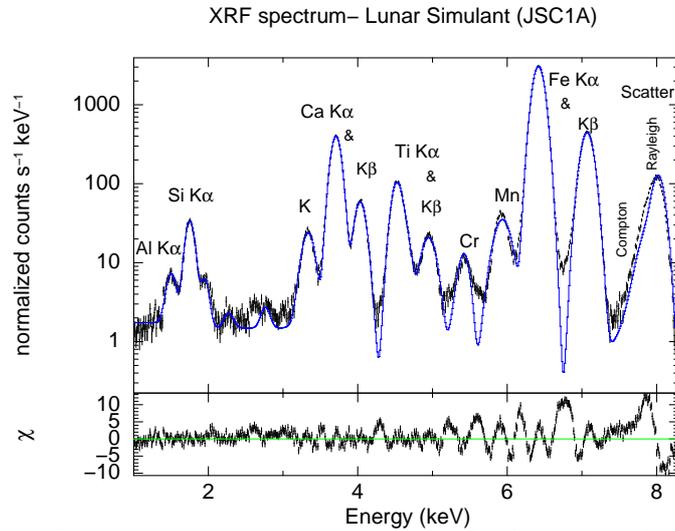}
		}
\hspace{4pt}%
		\subfigure[XRF spectrum - Lunar simulant (JSC-1A) - Fit with trace
		elements]{
		\includegraphics[width=7.05cm,angle=-90]
		{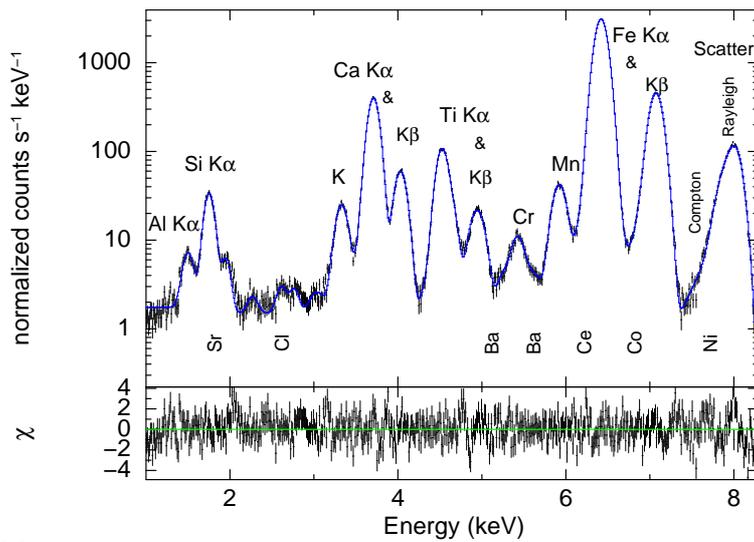}
		}\\
		\caption {Best fit for the observed XRF spectra of JSC-1A
		showing the lines of major elements present in the sample (a)
		without trace elements (b) with trace elements marked vertically.
		Residuals of fit ie., difference between data and model represented
		in terms of  standard deviation 1${\sigma}$ error bars (ref. delchi -
		XSPEC manual) are
		shown in the bottom panel of each figure.  Data points are shown
		with error bars; line shows the spectral model fit to the data. }
\label{specfitsjsc}
\end{figure}

\subfiglabelskip=0pt
\begin{figure}
	\centering
		\subfigure[XRF spectrum - Sittampundi Rock - Fit with trace
		elements]{
		\includegraphics[width=7.05cm,angle=-90]
		{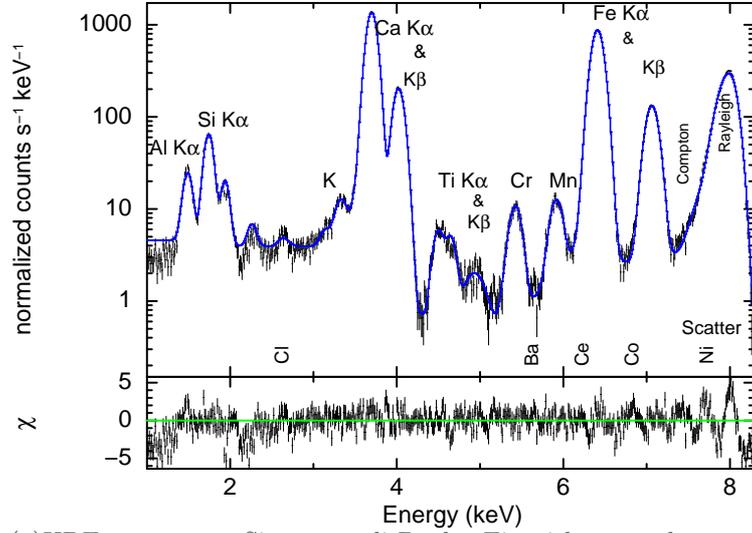}
		}
	\caption {Best fit for the observed XRF spectra of Sittampundi rock
	sample (anorthositic composition)
		showing the lines of major elements present in the sample.
		Prominent trace elements and other components are marked.
		Residuals of fit represented in terms of sigma (1${\sigma}$) error bars
		(ref. delchi - XSPEC manual) are shown in the bottom panel of each
		figure.}
\label{specfitsano}
\end{figure}

\begin{figure}%
		\centering
		\subfigure[Lunar simulant (JSC-1A) - representing lunar mare]{
		\includegraphics[width=11.05cm]
		{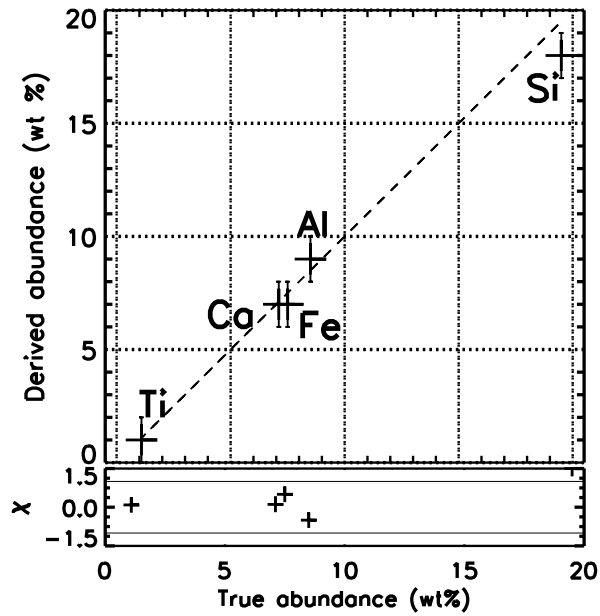}
		}
		\hspace{8pt}
		\subfigure[Sittampundi sample - representing lunar highland]{
		\includegraphics[width=11.05cm]
		{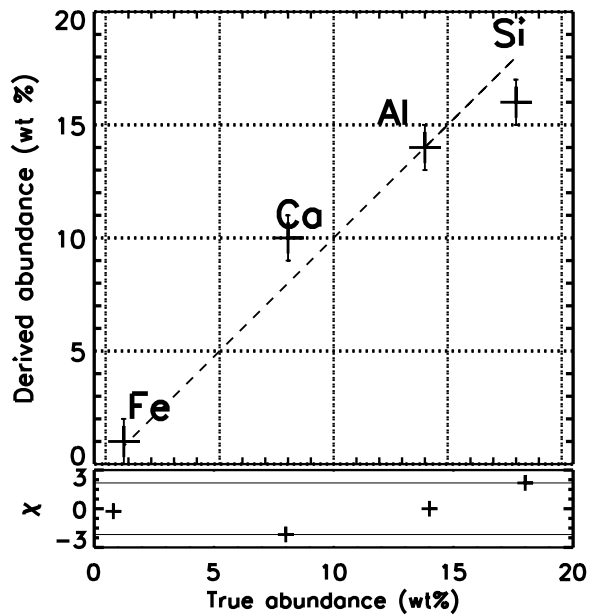}
		}\\
		\caption{Comparison plot between the derived abundance (from
		{\it x2abundance}) and true abundance (EDX measurements) of laboratory
		lunar analogue samples.	Residuals are plotted with 2${\sigma}$ confidence
		in the bottom panel}
		\label{labcompplots}
	\end{figure}

	\begin{table}[h]
	\caption{Comparison of weight percentage (wt\%) of major elements
	of JSC-1A as derived from {\it x2abundance} and EDX measurement values.}
	\label{JSC1Aabundance}
	\begin{tabular}{|l|l|l|l|}
	\hline
	\textbf{Element}&\textbf{Wt.\% from}&\textbf{Wt.\% from }&\textbf{Wt.\% from}\\
	\textbf{}&\textbf{EDX}&\textbf{x2abundance}&\textbf{x2abundance}\\
	& \textbf{measurement}&\textbf{(with trace}&\textbf{(without trace} \\
	& &\textbf{elements$)$}&\textbf{elements$)$} \\
	\hline
	Fe& 7.5 & 7 ${\pm}$ 1 & 7 ${\pm}$ 1 \\
	Ti& 1.1 & 1 $+$ 1 & 1 + 1 \\
	Ca& 7.1 & 7 ${\pm}$ 1 & 7 ${\pm}$ 1 \\
	Si&  19.5 & 18 ${\pm}$ 1 &17 ${\pm}$ 1 \\
	Al& 7.0 & 9 ${\pm}$ 1 &10 ${\pm}$ 1 \\
	\hline
	\end{tabular}
\end{table}

\begin{figure}%
		\centering
		\includegraphics[width=11.05cm]
		{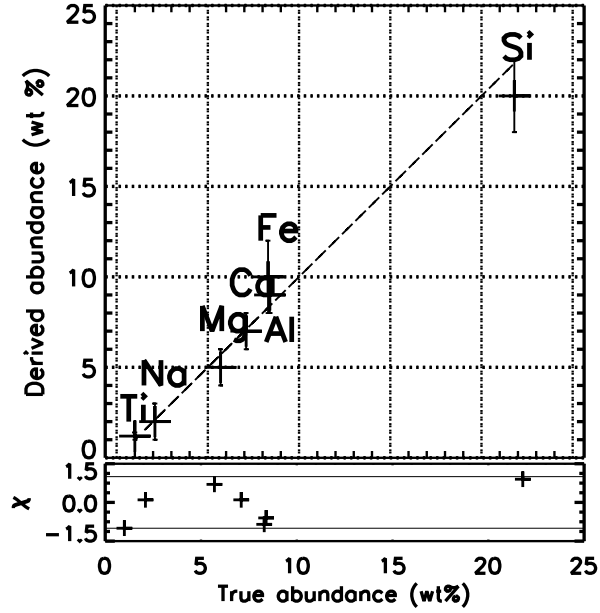}
		\caption{Comparison of abundance derived from simulated spectrum
		using x2abundance vs abundance assumed to create simulated spectrum
		in GEANT4 for the sample JSC-1A. Residuals are shown in the bottom panel.}
		\label{geantcompplots}

	\end{figure}

\begin{table}[h]
	\caption{Samples used in GEANT4 simulations}
	\label{geantsamples}
	\begin{tabular}{|c|l|l|}
	\hline
	\textbf{S.No.}&\textbf{Name}&\textbf{Major elements}\\
	\hline
	1& JSC-1A & Fe, Ti, Ca, Si, Al, Mg, Na, O \\
	2& Sittampundi sample & Fe, Ca, Si, Al, Na, O\\
	\hline
	\end{tabular}
\end{table}

\begin{figure}%
		\centering
		\includegraphics[width=11.05cm]
		{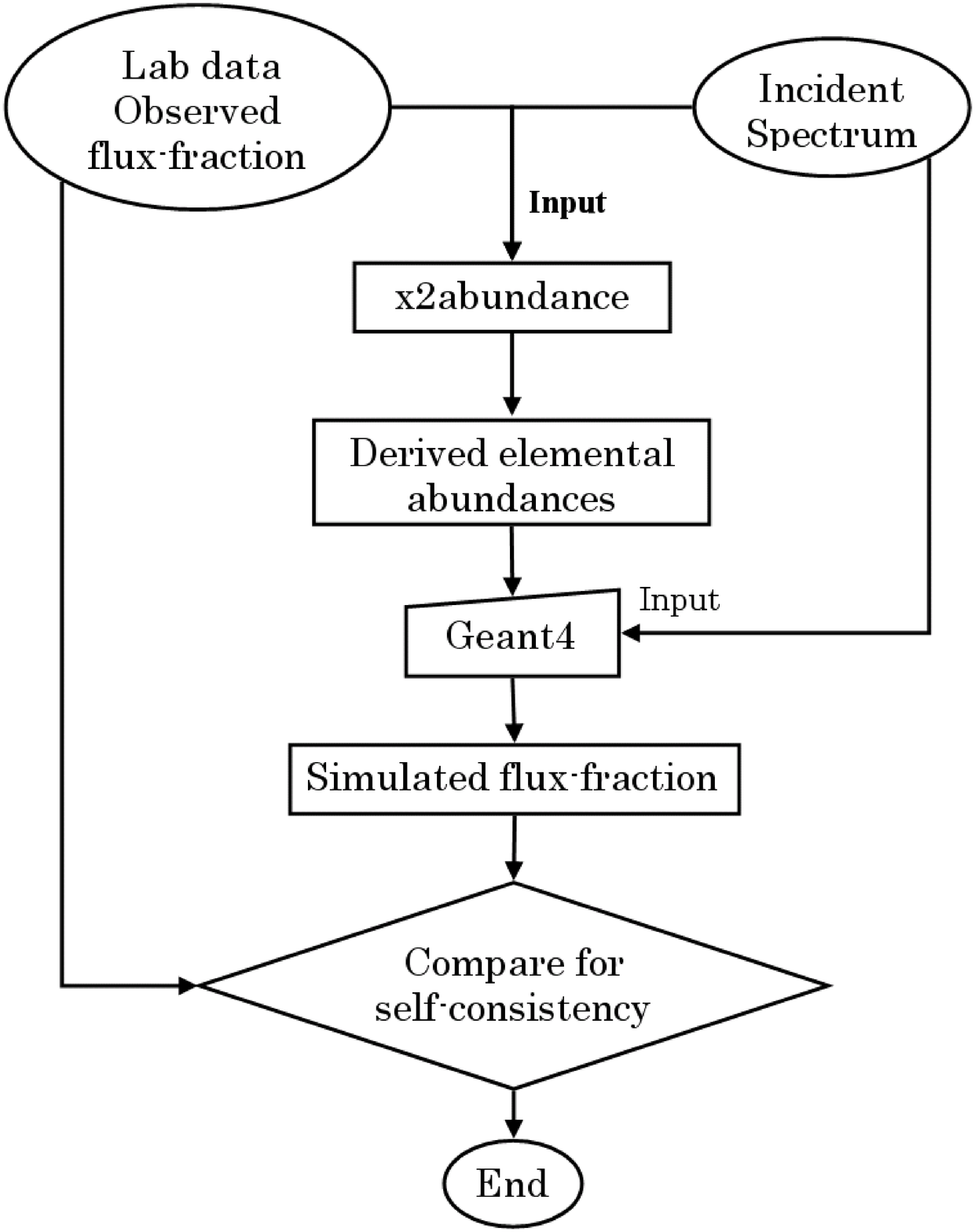}
		\caption{Steps involved in the cross-validation of {\it x2abundance}.}
		\label{closure}

	\end{figure}

\begin{figure}%
		\centering
		\includegraphics[width=10.05cm,angle=0]
		{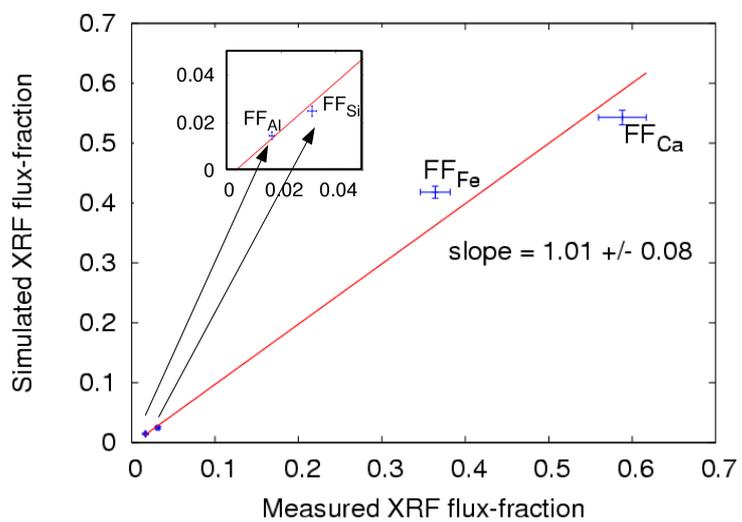}
		\caption{Comparison of XRF flux-fraction -  measured vs
		simulated - Sittampundi sample}
		\label{closure_ff}

	\end{figure}}

\end{document}